\documentclass[reprint,
twocolumn,
english,
aps,
prb,
superscriptaddress,
bibnotes,
amsmath,
amssymb,
floatfix,
longbibliography,
citeautoscript,
]{revtex4-2}

\usepackage[utf8]{inputenc}
\usepackage[T1]{fontenc}
\usepackage[colorlinks=true, allcolors=blue]{hyperref}
\usepackage{graphicx, dcolumn, siunitx, orcidlink}
\usepackage{threeparttable}
\usepackage{makecell}

\graphicspath{{figures/}}

\newcommand{\chitwo}{$\chi^{(2)} $ }
\newcommand{\SiN}[0]{Si$_3$N$_4$}

\begin{document}
\title{Ultrabroadband Milliwatt-Level Resonant Frequency Doubling on a Chip}

\author{Marco Clementi\orcidlink{0000-0003-4034-4337}}
\email{marco.clementi01@unipv.it}
\affiliation{Photonic Systems Laboratory, \'{E}cole Polytechnique F\'{e}d\'{e}rale de Lausanne,  1015 Lausanne, Switzerland}
\affiliation{Now at: Dipartimento di Fisica “A. Volta”, Università di Pavia, Via A. Bassi 6, 27100 Pavia, Italy}

\author{Luca Zatti\orcidlink{0000-0002-1280-9400}}
\affiliation{Dipartimento di Fisica “A. Volta”, Università di Pavia, Via A. Bassi 6, 27100 Pavia, Italy}

\author{Ji Zhou\orcidlink{0000-0001-8044-4426}}
\affiliation{Photonic Systems Laboratory, \'{E}cole Polytechnique F\'{e}d\'{e}rale de Lausanne,  1015 Lausanne, Switzerland}

\author{Marco Liscidini\orcidlink{0000-0003-4001-9569}}
\affiliation{Dipartimento di Fisica “A. Volta”, Università di Pavia, Via A. Bassi 6, 27100 Pavia, Italy}

\author{Camille-Sophie Brès\orcidlink{0000-0003-2804-1675}}
\affiliation{Photonic Systems Laboratory, \'{E}cole Polytechnique F\'{e}d\'{e}rale de Lausanne,  1015 Lausanne, Switzerland}

\date{\today}
\begin{abstract}
\noindent 
Microresonators are powerful tools to enhance the efficiency of second-order nonlinear optical processes, such as second-harmonic generation, which can coherently bridge octave-spaced spectral bands.
However, dispersion constraints such as phase-matching and doubly resonant conditions have so far limited demonstrations to narrowband operation.
In this work, we overcome these limitations showing ultrabroadband resonant frequency doubling in a novel integrated device, wherein the resonant enhancement of pump and second harmonic are individually addressed in two distinct and linearly uncoupled microring resonators, each adjusted to target the respective spectral band.
The two microresonators are designed and tuned independently, yet share a common interaction region that grants nonlinear coupling over a quasi-phase-matching bandwidth exceeding 200~nm, enabled by the inscription of a photoinduced $\chi^{(2)}$ grating.
The system allows to not only conveniently disentangle the design parameters of the two microresonators but also to reconfigure the doubly resonant condition electrically, and the phase-matching condition optically.
We demonstrate milliwatt-level addressable second-harmonic generation over the entire telecom band and then configure the device to internally generate and upconvert a Kerr frequency comb with bandwidth exceeding 100~nm and upconverted power up to 10~mW.
\end{abstract}

\maketitle

\begin{figure*}[ht!]
    \centering
    \includegraphics{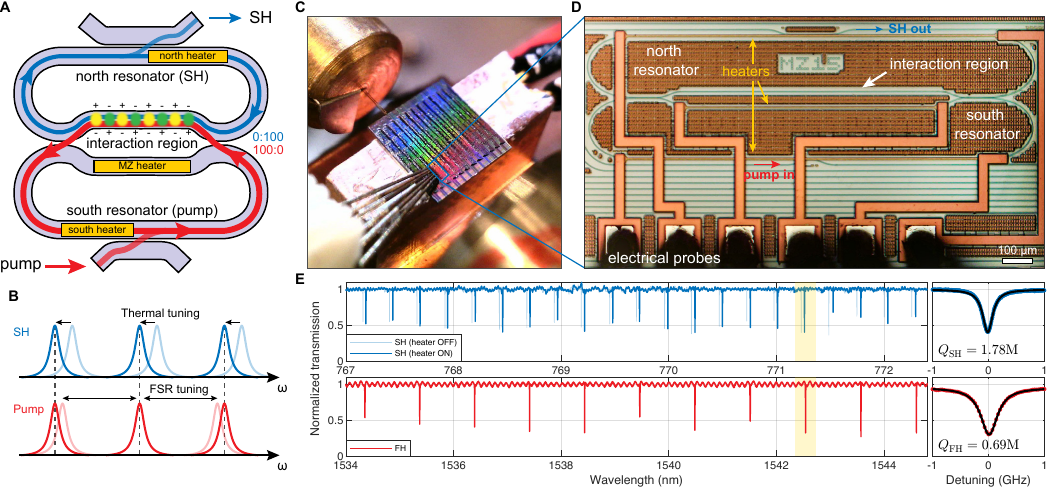}
    \caption{
        \textbf{Reconfigurable frequency doubling device.}
        \textbf{a.} Conceptual schematic of the device.
        The pump and SH circulate respectively in the south and north loops, sharing a fraction of the optical path where nonlinear interactions (AOP and SHG) occur.
        \textbf{b.} Resonances distribution for the north (\textit{top}) and south (\textit{bottom}) resonators, respectively at the pump and SH frequencies.
        The absolute resonance frequencies in the north (south) ring can be tuned by acting on the north (south) heater.
        The resonators' FSR can be individually tuned by tailoring the respective loop length.
        \textbf{c.} Micrographs of the fabricated chip and \textbf{d.} a single device.
        \textbf{e.} Low power transmission spectra for the TE$_{00}$ mode in the SH (\textit{top}) and pump (\textit{bottom}) bands.
        The detail shows a typical pair of resonances involved in the SHG process and associated Lorentz fit.
    }
    \label{fig:fig1}
\end{figure*}

\section*{Introduction}

\noindent Second-order nonlinear optical processes, such as second-harmonic generation (SHG), are fundamental tools in photonics technology, as they enable converting light across widely separated regions of the optical spectrum while preserving its coherence \cite{Boyd2020}.
They find widespread application in the engineering of light sources \cite{svelto2010principles}, imaging \cite{Campagnola2011}, ultrafast physics \cite{Trebino2000}, material science \cite{shen1989}, and quantum technology \cite{Kwiat1995}, to name a few.
Leveraging the nonlinear dielectric response of a medium typically requires high field intensities, which nowadays can be effectively achieved through high peak power pulsed sources in bulk systems --- or  by exploiting spatial and temporal light confinement in integrated structures.

Microring resonators stand, among various microcavity geometries proposed \cite{Furst2010, Wang2020}, as a prime choice for efficient on-chip frequency doubling,  thanks to the possibility to design devices with high Q resonances at both fundamental- (FH) and second-harmonic (SH).
Highly efficient frequency doubling has notably been achieved in  \chitwo material platforms such as thin-film lithium niobate (TFLN), with conversion efficiency ($\mathrm{CE}=P_{\rm SH}/P_{\rm FH}^2$, $P_{\rm FH}$ and $P_{\rm SH}$ being respectively the pump and SH power) as high as 5,000,000\%/W \cite{Lu2020},  III-V materials (e.g. AlN \cite{Bruch2018}, GaN \cite{Wang2020}, GaAs \cite{Kuo2014}, GaP \cite{Lake2016}), and silicon carbide (SiC) \cite{Lukin2020}.
Despite remarkable advances, these materials generally still suffer from limited power handling capability, a non-standardized fabrication process, and material-specific limitations.

An alternative approach is represented by silicon nitride (Si$_3$N$_4 $) integrated photonics \cite{Liu2021}, which combines low loss, a well-established and CMOS-compatible fabrication, and excellent power handling capabilities.
Although lacking intrinsic \chitwo due to its amorphous nature, Si$_3$N$_4 $ can be endowed with a photoinduced second-order nonlinearity through the coherent photogalvanic effect
\cite{Billat2017, Porcel2017, Hickstein2019, Yakar2022}. 
In resonant structures, this can yield high CE (up to 2,500\%/W) \cite{Lu2021} and record-high generated power (>10~mW) \cite{Nitiss2022}.
These advances have made it possible to realize complex chip-scale functionalities such as frequency comb self-referencing \cite{Hickstein2019, Nitiss2020broadband} and self-injection-locked SHG \cite{Clementi2023achipscale, Li2023}.

While microring resonators are capable of highly efficient frequency doubling, their practical usage is hindered by limitations in bandwidth and poor wavelength tunability.
Indeed, broadband SHG in a ring resonator must satisfy three requirements simultaneously: (i) phase-matching, (ii) doubly resonant condition for pump and SH, and (iii) free spectral range (FSR) matching.
Perfect phase-matching is usually prevented by material and modal dispersion, but it can be obtained through intermodal phase-matching \cite{Lu2021} or through quasi-phase-matching (QPM) --- the latter either exploiting the natural birefringence of the medium ($\overline{4}$-QPM) \cite{Kuo2014} or the electric-field poling of a ferroelectric waveguide \cite{Lu2020}.
Doubly resonant high-Q microrings are also typically unachievable deterministically, even by tailored designs, due to fabrication tolerances, for the resonance position must be controlled with a precision comparable with the linewidth.
A common workaround is thermal tuning \cite{Nitiss2023}, where the slightly different thermo-optic effect on pump and SH resonances can compensate for small deviations from a nearly doubly resonant condition statistically obtained among many fabricated samples.
However, this method is not scalable, does not allow for the precise control of the operation wavelength, and becomes unpractical in systems with large FSR.
If a doubly resonant condition is difficult to achieve at a given pump wavelength, broadband SHG involving multiple resonances is even more challenging.
Indeed, the FSRs at pump and SH frequencies are in general different: therefore, high-Q resonators only enable doubly resonant SHG over a small number of resonances at the time, restricting applications such as frequency comb upconversion to a limited number of spectral lines \cite{Miller2014, Hu2022} and, more generally, setting a trade-off between bandwidth and CE \cite{Guo2018, He2019, Wilson2020}.
In principle, this can be mitigated by adjusting the waveguide cross-section to match the group velocity \cite{Nitiss2020broadband}, but at the cost of dispersion engineering, which is already required for controlling the comb properties or generating soliton states.

In this work, we address all of these issues, inherent to the single-ring approach, by demonstrating a device where the pump and SH fields are associated with two separate resonators that yet share a common path where nonlinear processes can occur. 
This change of paradigm allows one to adjust the optical properties of the system in different spectral regions independently and achieve efficient nonlinear interaction without dispersion engineering \cite{Menotti2019, Tan2020, Sabattoli2021, Sabattoli2022, Zatti2022, Zatti2023pra}. 
First, a photoinduced nonlinearity is realized in the Si$_3$N$_4$ coupling region where pump and SH coexist, whose periodicity automatically satisfies the QPM condition over a broad bandwidth (all-optical poling process, AOP \cite{Nitiss2022}).
Second, the families of azimuthal modes of the two resonators can be independently tuned through electric actuators to make sure that a doubly resonant condition is always satisfied.
Third, the microresonator lengths are tailored to grant FSR matching at the design level. 
Finally, efficient in-coupling and extraction of light at the pump and SH wavelength are obtained by optimizing the couplers to the two resonators independently.
We investigate the device operating principle by two-photon microscopy, showing that the AOP process occurs in the region where the circulating pump and SH field interact, proving the linearly uncoupled nature of the two resonators.
We then showcase the potential of this approach by reconfiguring the device to achieve SHG with milliwatt-level generated power over the C and L telecom bands, limited only by the bandwidth of our amplifiers.
Finally, we demonstrate the broadband operation of the device by using it to generate a modulation instability comb that we frequency-double intracavity, within the same structure, over a pump bandwidth of nearly \SI{100}{\nano\meter} (\SI{12.5}{\tera\hertz}).

\begin{figure*}[ht!]
    \centering
    \includegraphics{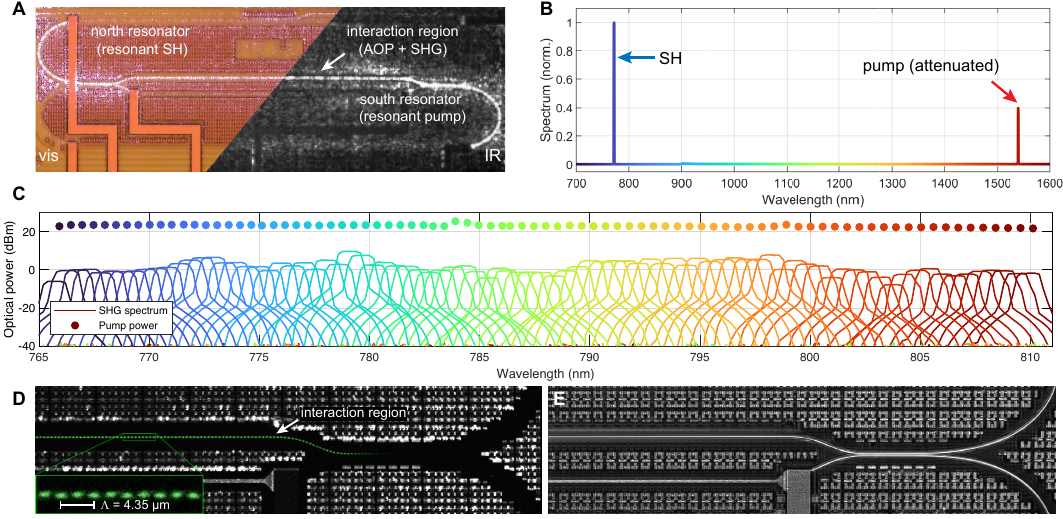}
    \caption{
    \textbf{All-optical poling and second-harmonic generation.}
    \textbf{a.} Micrograph of a typical device in frequency doubling operation.
    The visible camera (\textit{left}) emphasizes the resonant field enhancement of the generated SH in the north loop, while the infrared (IR) camera (\textit{right}) shows the field enhancement in the south loop.
    Note that the two fields overlap in the shared path, where AOP and simultaneous SHG occur.
    \textbf{b.} Typical experimental SHG spectrum, showing generated light at \SI{770}{\nano\meter} (pump attenuation: 27.5~dB).
    \textbf{c.} Collection of SHG spectra associated with pumping in the C and L bands (resolution: \SI{1}{\nano\meter}).
    Each trace is a detail of a spectrum as shown in panel b.
    The heater currents are slightly adjusted for each data point to optimize the generated power.
    \textbf{d.} Two-photon and \textbf{e.} one-photon micrograph of one of the devices after inscription of the nonlinear QPM grating through AOP.
    TPM imaging reveals a periodic $\lvert\chi^{(2)}\rvert^2$ distribution that only exists in the interaction region, confirming the linearly uncoupled nature of the process.
    }
    \label{fig:fig2}
\end{figure*}

\section*{Results}

\subsection*{Device description and operating principle}
\noindent A conceptual schematic of the reconfigurable SHG source is shown in Fig.~\ref{fig:fig1}a.
The device is composed of two “racetrack” resonators designed to operate respectively in the pump (“south”) and SH (“north”) spectral bands.
Both resonators are nearly critically coupled to their bus waveguide  for the fundamental transverse-electric mode (TE$_{00}$).
A Mach-Zehnder interferometer (MZI) structure is used to introduce an interaction region between the two resonators by using two directional couplers designed to operate near 100:0 splitting ratio in the pump band and near 0:100 in the SH band \cite{Zatti2023pra}. 
Finally, a thermo-optic phase shifter (“heater”) is placed on each resonator and on the lower arm of the MZI, allowing, respectively, to tune the set of resonant modes at pump/SH and to compensate for any residual linear coupling between the two racetracks.

The resulting resonances distribution is sketched in Fig.~\ref{fig:fig1}b: differently from the case of a single resonator, the two sets of modes can be independently shifted to achieve a doubly resonant condition by acting on the north/south heater.
Moreover, by engineering the length of the two resonators, one can by-design match the FSRs of the two rings in order to allow simultaneous frequency doubling on multiple resonances, according to the formula:
\begin{equation}
    \mathrm{FSR_{\mathrm{FH}}} =
    \frac{v_\mathrm{g}^\mathrm{FH}}{L_\mathrm{FH}} =
    \frac{v_\mathrm{g}^\mathrm{SH}}{L_\mathrm{SH}} =
    \mathrm{FSR_{\mathrm{SH}}}
    \label{eq:fsr}
\end{equation}
where $v_\mathrm{g}^\mathrm{FH(SH)}$ is the group velocity of the pump (SH) mode and $L_\mathrm{FH(SH)}$ is the optical path length of the south (north) resonator.
Note that in the case of a single resonator (where $L_\mathrm{FH}=L_\mathrm{SH}$), FSR matching can only be achieved through group velocity matching, posing a significant constraint on dispersion engineering.

A typical fabricated device is shown in Figs.~\ref{fig:fig1}c-d. 
We use a lensed fiber to couple light to the chip and electrical probes to actuate the heaters (see Methods).
Typical transmission spectra in the pump and SH bands are shown in Fig.~\ref{fig:fig1}e.
By acting on the south and north resonator heaters we are able to observe a rigid and independent shift of the respective resonance spectra (the north heater case is shown in Fig.~\ref{fig:fig1}e).
Similarly, acting on the MZI heater enables us to dynamically vary the pump resonances visibility and linewidth, as a result of power leakage to the north resonator, strongly overcoupled in the pump band  (see Supplementary Note 1).
Once these values are optimized, from best fit of the Lorentzian lines recorded we estimate a loaded Q factor of around $0.69\times10^6$ at the pump and $1.78\times10^6$ at the SH.

\begin{figure*}[ht!]
    \centering
    \includegraphics{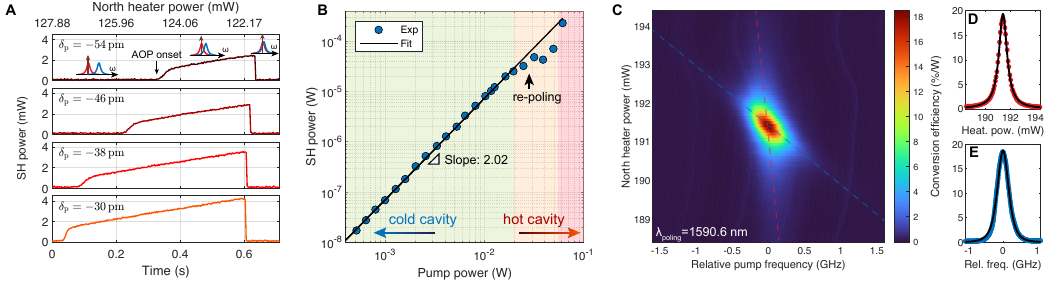}
    \caption{
    \textbf{Reconfiguration of the nonlinear interaction.}
    \textbf{a.} SHG traces obtained by ramping down the north heater power (i.e. slowly blue-shifting the north ring SH resonance).
    Each trace corresponds to a different detuning $\delta_\mathrm{p}=\lambda_\mathrm{res}-\lambda_\mathrm{pump}$ of the pump laser from the hot cavity resonance at its maximal shift.
    The asymmetric lineshape is attributed to the presence of thermal shift in the north resonator.
    \textbf{b.} SHG power scaling measurement and quadratic fit.
    \textbf{c.} High resolution SHG map at in the cold cavity regime after inscription of the nonlinear grating.
    Here the pump power is kept low (\SI{0.2}{\milli\watt}) to prevent thermal shifts as well as the triggering the AOP process.
    The red and blue dashed lines highlight respectively the pump and SH resonance peaks, which red-shift for increasing north heater power.
    The peak values are plotted in panels \textbf{d} and \textbf{e} and fitted with a Lorentzian and squared Lorentzian function respectively (black solid line).
    }
    \label{fig:fig3}
\end{figure*}

\subsection*{Addressable doubly resonant SHG}
\noindent
To investigate SHG, we pump the south resonator with an amplified tunable laser and configure the device with the procedure described in the Methods.
The occurrence of the AOP process is marked by a sudden increase of the SHG power from the north resonator.
Visible and infrared imaging of the device in operation (Fig.~\ref{fig:fig2}a) provides straightforward evidence of the linearly uncoupled nature of the process. 
Light scattering from the north and south resonators during the SHG process, when acquired with either a visible or an infrared camera, suggests that the SH and pump fields are circulating only in their respective resonators.
A typical spectrum of the output light is also shown in Fig.~\ref{fig:fig2}b, confirming the SHG process and the absence of spurious spectral components.
By repeating this procedure, we reproduce a similar result among several devices with different FSR and coupling conditions, systematically recording a generated SH power of up to \SI{10}{\milli\watt} (with a pump power of \SI{220}{\milli\watt}).

We then focus our attention on a single device, characterized by a nominal FSR of about \SI{130}{\giga\hertz}, and investigate its suitability for efficient SHG across multiple resonances.
To this end, we tune the pump wavelength to each of the resonances within the C and L telecom bands (1530-1620~nm) and repeat the optimization procedure, each time reconfiguring the QPM grating by AOP.
Fig.~\ref{fig:fig2}c shows the recorded SHG spectra, all corresponding to an output power in the milliwatt level (average: 1.4~dBm, standard deviation: 2.4~dB, see Supplementary Figure 6 for more details), proving the capability of the integrated nonlinear device to operate across a wideband range when electrically reconfigured.
Note that this demonstration is limited by the availability of amplifiers, while the nonlinear device is expected to be capable of covering an even broader spectrum, limited only by the bandwidth of the resonators' point couplers and of the MZI, estimated at about \SI{150}{\nano\meter} (\SI{19}{\tera\hertz}).

Further insight on the AOP process can be gathered by two-photon microscope (TPM) imaging of the inscribed \chitwo grating (Fig.~\ref{fig:fig2}d) \cite{Hickstein2019, Nitiss2020formation}.
Compared with regular (one-photon) micrographs (Fig.~\ref{fig:fig2}e), TPM images clearly show that the nonlinear grating inscription occurs only in the upper arm of the interferometer, corroborating the hypothesis of linear uncoupling. 
The poling period is measured to be around \SI{4.35}{\micro\meter}, in perfect agreement with the simulated values of effective indices for the TE$_{00}$/TE$_{00}$ mode pair at pump/SH.

\begin{figure*}[ht!]
    \centering
    \includegraphics{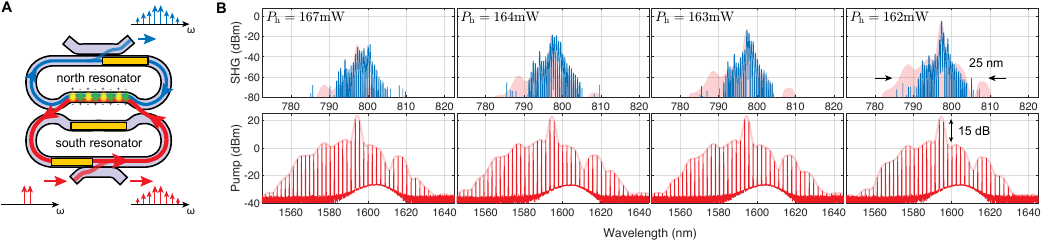}
    \caption{
    \textbf{Stimulated FWM comb upconversion.}
    \textbf{a.} Schematic of the stimulated FWM comb generation.
    The south loop is fed with a bichromatic pump created by mixing and amplifying two tunable sources spaced at a multiple of the FSR.
    The resonantly enhanced $\chi^{(3)}$ interaction in the south resonator produces an incoherent comb through cascaded FWM.
    The comb is then frequency doubled in the interaction region, where it participates to AOP and related SHG and SFG, and finally output from the north resonator coupler.
    \textbf{b.} 
    Generated comb (spacing: 2 FSR) for varying resonance detuning configurations.
    When red-detuning the north ring resonances from the optimal condition, the entire SHG comb power is gradually decreased. 
    The shaded area in the top graphs represents the squared FH comb envelope (shaded trace in the bottom graphs) rescaled to the maximum of the generated SH, for comparison.
    }
    \label{fig:fig4}
\end{figure*}

\begin{figure*}[ht!]
    \centering
    \includegraphics{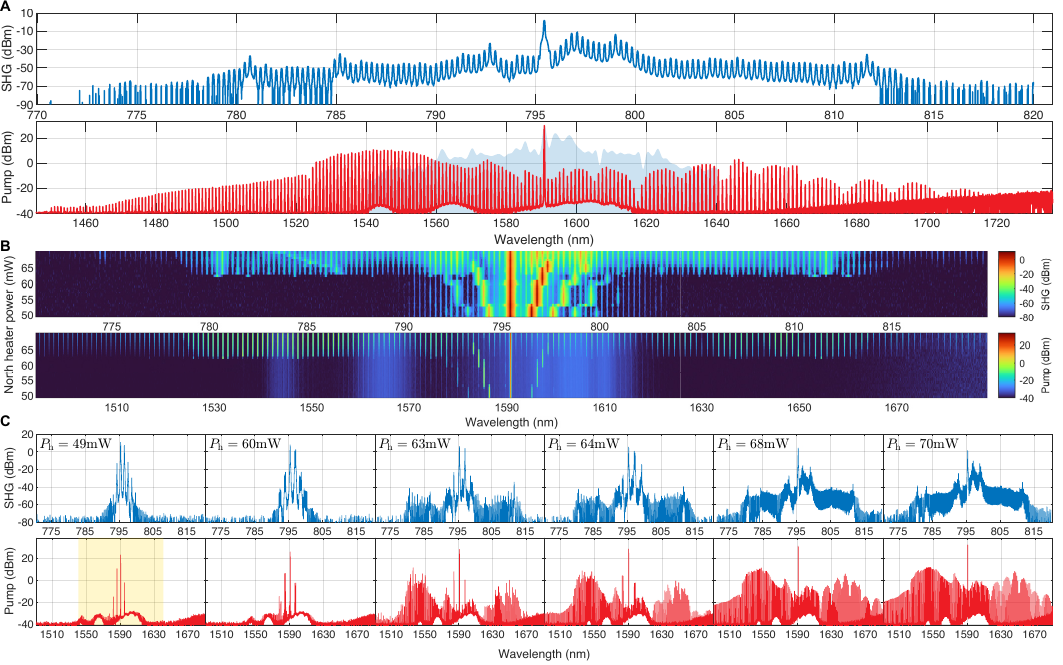}
    \caption{
    \textbf{Modulation instability comb generation and upconversion.}
    \textbf{a.} Generated MI comb (\textit{bottom}) and its upconverted spectrum (\textit{top}), with a pump power of \SI{1.6}{\watt}.
    The shaded area corresponds to the upconverted spectrum extent.
    \textbf{b.} Accumulated optical spectra of FH (\textit{bottom}) and SH (\textit{top}) for varying north heater power.
    A crossover from Turing patterns to the broadband MI comb can be observed in combination with a decrease of the overall generated power.
    \textbf{c.} Selected spectra from panel \textbf{b}.
    }
    \label{fig:fig5}
\end{figure*}

\subsection*{Nonlinear properties}
\noindent
To gather insight on the nonlinear properties of the developed device, we first perform a broadband mapping of the AOP process \cite{Nitiss2023, Clementi2023achipscale}, as detailed in Supplementary Note 2.
We then focus our attention on a single resonance (around \SI{1590}{\nano\meter}) and investigate the AOP dynamics at different values of pump detuning (Fig.~\ref{fig:fig3}a).
We first set the pump wavelength close to resonance, and then slowly ramp down the north heater power, inducing a blue-shift of the SH resonance.
Consequently, the SH field enhancement is progressively increased until a sudden spike in the SH power is observed, marking the AOP onset, where the conditions for poling of the north resonator are met \cite{Nitiss2022}.
When further decreasing the north heater power, the generated SH increases owing to the increased field enhancement, until a sudden drop is observed.
We attribute this to the specific dynamics of the AOP process in resonant structures \cite{Zhou2025}.
We repeat the measurement while further tuning the pump close to resonance, observing an earlier occurrence of the AOP onset, an overall increase of the peak generated SH, and a systematic drop occurring at nearly the same SH detuning conditions.

The experiments described so far entail circulating pump and SH intensities sufficiently high to trigger the AOP process.
However, it is known that the \chitwo grating, once inscribed, can subsist for very long time, exceeding several months at room temperature \cite{Nitiss2020formation}.
We therefore investigate the nonlinear response of the system in a regime of low pump power, insufficient to trigger the erasure or reconfiguration of the grating, after previous poling a selected resonance.
To establish the limits of such “cold cavity” regime, we perform a scaling analysis of the generated SH power as a function of the pump power in the bus waveguide, as shown in Fig.~\ref{fig:fig3}b.
For each value of the pump power, the wavelength is slowly scanned through the resonance from the blue to the red side, and the peak SHG power is recorded.
The resulting trend shows a quadratic increase of the generated SH for pump powers below approximately \SI{20}{\milli\watt}.
When the pump power is further increased, a deviation from such trend is recorded: we attribute this to the reconfiguration of the \chitwo grating \cite{Zhou2025}. 
When further increasing the pump power, the generated SH restores the quadratic trend, reentering the combined AOP/SHG process (“hot cavity” regime).
The cold- and hot-cavity regimes are therefore desirable for efficient SHG, while the intermediate regime of reconfiguration should be avoided.
At higher power, we expect the quadratic trend to eventually saturate owing to the increased nonlinear photoconductivity \cite{Nitiss2022, Yakar2022}.

With these results in hand, we investigate the SHG process in the cold cavity regime, for which we map the CE as a function of the pump wavelength and north heater power after poling (Fig.~\ref{fig:fig3}c).
To minimize the impact of thermal bistability on the measurement, here we keep the pump power as low as \SI{0.2}{\milli\watt}.
The result displays two main branches: the first (red dashed line), nearly vertical, represents the pump resonance, which doesn't undergo a shift when increasing the north heater power (except for a slight thermo-optic crosstalk); the second (blue dashed line), oblique, is associated with the SH resonance, which experiences the most important shift rate.
In Figs.~\ref{fig:fig3}d-e we show the slices obtained along these trends, which we fit with a Lorentzian and squared-Lorentzian model respectively.

\subsection*{Frequency comb upconversion}
\noindent
To showcase the potential of our device, we investigate its suitability for upconversion of a frequency comb.
We first inject in the south resonator two separate L band CW lasers, amplified and tuned to match an integer multiple of the resonator's FSR, as sketched in Fig.~\ref{fig:fig4}a, and tune the south resonator heater in order to couple efficiently to the respective resonances.
Enhanced by the south resonator, the circulating field produces a cascade of stimulated four-wave mixing (FWM) processes, resulting in the formation of an incoherent frequency comb, whose bandwidth is determined by the pump line spacing, detuning, and waveguide dispersion.

The circulating FWM comb can trigger the AOP process.
To this end, we slowly decrease the north heater power while monitoring the generated spectrum at both pump and SH with two optical spectrum analyzers (OSA), as shown in Fig.~\ref{fig:fig4}b.
As soon as the appropriate detuning conditions are met, the north resonator is observed to scatter near-visible light, confirming the onset of the AOP process.
The SHG spectrum displays a set of equally spaced comb lines with the same frequency spacing (1/4 wavelength spacing) as the pump, and upconverted power in the milliwatt level per line.
We interpret the observed spectrum as the result of a combination of SHG and SFG processes, all phase matched by the same \chitwo grating, made possible by the matching of the two resonators' FSRs, which do not affect the pump spectrum.

The envelope of the frequency doubled comb systematically replicates that of the pump spectrum squared (red shaded area), as expected, with a small deviation that we attribute to the residual FSR mismatch and to the effect of GVD.
Note that, in contrast, the QPM bandwidth (estimated around \SI{202}{\nano\meter}) does not limit the generation bandwidth, owing to the short length of the grating  $L_\mathrm{MZI}\approx \SI{361}{\micro\meter}$ (see Methods, Eq.~\ref{eq:SHG_power}).
By varying the comb lines spacing to integer multiples of the FSR, it is also possible to broaden the fundamental comb (Supplementary Note~3), while the frequency doubling bandwidth remains nearly constant.

\subsection*{Simultaneous comb generation and upconversion}
We then investigate the system suitability to simultaneously generate and upconvert a Kerr comb.
To this end, we use a single high-power (1.6 W) line around \SI{1591}{\nano\meter} to pump the south resonator and produce a modulation instability (MI) comb (Fig.~\ref{fig:fig5}a).
To trigger the comb and maximize its extent, we deliberately introduce a weak linear coupling between the north and south resonators (Supplementary Note 4), which produces an anticrossing-type perturbation in the otherwise normal GVD profile \cite{Xue2016} and simultaneously broadens the linewidth of higher-frequency resonances, until we establish an optimal span of around \SI{300}{\nano\meter}.
By adjusting the north heater power to reach a doubly resonant condition, an upconverted comb spectrum spanning nearly \SI{50}{\nano\meter} is observed in the SH band, corresponding to a \SI{100}{\nano\meter} extent in the pump band. 
The upconverted comb spectrum replicates the envelope of the pump comb, but compared to the FWM cases, the pump comb shape also depends on the north heater power, as shown in (Figs.~\ref{fig:fig5}b-c).
We see that a lower heater power (lower SH detuning, while keeping unchanged the excitation conditions) results in the most efficient SHG, with upconverted power exceeding \SI{10}{\milli\watt} per line (Figs.~\ref{fig:fig5}c, first panel), but at the same time a generated comb spectrum of only few lines, characteristic of primary combs.
When increasing the heater power, hence  increasing the SH detuning, we observe a gradual transition towards the full MI comb formation at the pump band and broader upconversion, albeit paired with a decrease in the overall upconverted power.
This phenomenological trade-off trend can be attributed to pump depletion, which sets a competition between the $\chi^{(3)}$ process yielding the comb formation and the AOP/SHG process.

\section*{Discussion}
\begin{table*}[hbt!]
	\begin{threeparttable}
	\caption{Comparison of state-of-the-art integrated platforms for SHG at telecommunication wavelength.}
	\label{tab:tab1}
  \begin{center}
  \footnotesize\centering
  \begin{tabular}
  { m{4cm}<{\centering} 
  m{2.8cm}<{\centering} 
  m{2.8cm}<{\centering} 
  m{2.8cm}<{\centering} 
  m{4.7cm}<{\centering} }
        \hline
        \textbf{\makecell{Platform}} & 
        \textbf{\makecell{On-chip CE \\(\%/W)}} & 
        \textbf{\makecell{SHG bandwidth \\ (nm)}} &
        \textbf{\makecell{Max on-chip \\SH power (mW)}} &
        \textbf{\makecell{Phase-matching \\method}}\\ 
        \hline
        TFLN waveguide ($X$-cut) \cite{Chen2024}
        & 9,500 
        & 0.97
        & 16.5
        & QPM (electric field poling)
        \\
        TFLN microring ($Z$-cut) \cite{Lu2019Periodicallypoledthin}
        & 250,000
        & >10 
        & 0.1 
        & QPM (electric field poling)
        \\
        TFLN microring ($Z$-cut) \cite{Lu2020}
        & 5,000,000
        & 0.005$^\mathrm{a}$ 
        & 0.02 
        & QPM (electric field poling)
        \\
        TFLN microring ($Z$-cut) \cite{He2019}
        & n.a. 
        & 80$^\mathrm{b}$ 
        & n.a.
        & n.a.
        \\
        AlN microring \cite{Guo2016b}
        & 2,500 
        & 0.42 
        & 3.3 
        & Intermodal
        \\
        AlN microring \cite{Bruch2018}
        & 17,000
        & 5  
        & 11
        & Intermodal
        \\   
        AlN microring \cite{Guo2018}
        & 7 
        & 60$^\mathrm{b}$     
        & 0.61                 
        & Intermodal
        \\   
        GaP microdisk \cite{Lake2016}
        & 38
        & 1
        & $4\times 10^{-5} $
        & $ \overline{4}$-QPM
        \\
        GaP microring \cite{Wilson2020}
        & n.a.
        & 40$^\mathrm{b}$ 
        & n.a. 
        & n.a.
        \\
        SiC microring \cite{Lukin2020}
        & 360
        & 0.03$^\mathrm{a}$
        & 0.001
        & Intermodal
        \\
        \SiN\ waveguide \cite{Billat2017}
        & 0.05
        & 4.5 
        & 0.16 
        & QPM (all-optical poling)
        \\
        \SiN\ waveguide \cite{Nitiss2020broadband}
        & 0.15
        & >25 
        & n.a.
        & QPM (all-optical poling)
        \\
        \SiN\ waveguide \cite{Hickstein2019}
        & 0.003
        & >100 
        & 0.6$^\mathrm{c}$ 
        & QPM (all-optical poling)
        \\
        \SiN\ microring \cite{Miller2014}   
        & 0.16
        & 10$^\mathrm{b}$
        & 1
        & Intermodal
        \\
        \SiN\ microring \cite{Lu2021}
        & 2,500
        & 0.06$^\mathrm{a,d}$
        & 2.2
        & Intermodal
        \\
        \SiN\ microring \cite{Nitiss2022}
        & 47.6
        & 0.6$^\mathrm{a,d}$
        & 12.5
        & QPM (all-optical poling)
        \\
        \SiN\ microring \cite{Nitiss2023}
        & 2
        & 2.3
        & 2
        & QPM (all-optical poling)
        \\
        \textbf{\makecell{\SiN\ microresonators \\ system (this work) }}         
        &\textbf{40}
        & \textbf{>100$^\mathrm{b}$}
        & \textbf{>10}
        & \textbf{QPM (all-optical poling)}
        \\
		\hline
		\end{tabular}
		\begin{tablenotes}
        \footnotesize
        \item[] 
        The bandwidth is estimated at 10 dB; power and CE are intended at CW regime unless otherwise stated.
        n.a.: data not available.
        \item[a] Doubly resonant SHG limited to a single resonance pair at pump/SH.
        \item[b] Estimated from the overall extent of the upconverted comb spectrum.
        \item[c] Average value at pulsed regime.
        \item[d] With thermal tuning of the resonance.
	\end{tablenotes}
    \end{center}
	\end{threeparttable}
\end{table*}

\noindent
The device developed represents the first demonstration of addressable doubly resonant SHG through external tuning, overcoming several fundamental limitations inherent in the single-resonator approach.
Our strategy can be generally set in the context of coupled resonator systems (or “photonic molecules”) which have been recently employed to improve the performance of other nonlinear processes, such as comb generation \cite{Helgason2023}, self-injection locking \cite{Ji2024}, and quantum light generation \cite{Zhang2021, Nigro2022}.
However, our approach differs fundamentally from the above due to the deliberate choice of eliminating the linear coupling, with the advantages of (i) strongly suppressing the parasitic loss due to overcoupling of north resonator's modes at the pump (ii) avoiding the formation of supermodes, associated with a lower field enhancement (equivalently, larger mode volume) and (iii) avoiding (or controlling) alterations to the dispersion profile associated with mode anti-crossings.
It's also worth noticing that our paradigm shares similarities with hybrid racetrack geometries recently employed for the demonstration of $\chi^{(2)}$ optical parametric oscillators in TFLN \cite{Ledezma2023, Stokowski2024}, where a non-resonant mode at short wavelength is linearly uncoupled, but nonlinearly coupled with the resonant modes of a microresonator at longer wavelengths.

Our strategy, originally introduced in the context of spontaneous parametric sources \cite{Menotti2019, Tan2020, Sabattoli2021, Sabattoli2022, Zatti2022} and here combined with AOP, enables electrical control over the doubly resonant condition within a pump bandwidth exceeding \SI{90}{\nano\meter} (\SI{11.2}{\tera\hertz}), only limited in this demonstration by the range of our amplifiers.
The system can systematically generate milliwatt-level powers, with CE as high as 40\%/W in CW regime.
These values are already useful for several practical applications and can likely be improved through further reduction of the propagation loss and by achieving a better rejection of the residual linear and intermodal coupling at the pump, further mitigating the need for high pump power.
Note that the CE could be further improved by increasing the resonator quality factors without compromising the phase-matching bandwidth, thanks to the flexible engineering of resonance positions. 
Alternative designs for linear uncoupling (see, e.g., Ref. \cite{Zatti2023pra}) could also help mitigate the current efficiency penalty, albeit at the expense of reduced QPM bandwidth (see Methods).

Moreover, our device enables the broadband upconversion of frequency combs through the engineering of a “synthetic” group velocity matching condition (Eq.~\ref{eq:fsr}), which we showcase over a pump bandwidth of about \SI{100}{\nano\meter} (\SI{12.5}{\tera\hertz}) and upconverted power exceeding \SI{10}{\milli\watt} per line, with potential applications to self-referencing of frequency combs.
While it is worth noting that few earlier demonstrations using the single resonator approach have already shown comparable upconversion bandwidth \cite{Guo2018, He2019, Wilson2020}, the generated powers reported so far have always been at least two orders of magnitude lower than in the present demonstration, and the nature of the modes involved at the SH has not always been clarified.

Table \ref{tab:tab1} provides an overview of several integrated platforms at the state of the art, showing the key figures of merit for broadband SHG.
While some demonstrations have reported higher values separately, our result effectively represents the best trade-off between CE, bandwidth, and generated power, especially among CMOS-compatible platforms.

A further perspective advantage of our approach is the possibility of continuously tuning the doubly-resonant SHG. 
By synchronizing a pump laser scan with the tuning of the south and north heaters, one could potentially achieve a “gap-free” tuning of the generated SH field, which could be particularly beneficial to address specific SH wavelengths, such as those associated with atomic transitions.
Essential requirements to this end are (i) the possibility to continuously tune the pump and SH resonances across at least one FSR (here around \SI{130}{\giga\hertz},  realistically achievable with standard heater modules) and (ii) a seamless transition between different pairs of resonances involved in the AOP-SHG process, that should be slow enough to allow the reconfiguration of the grating.

While the present demonstration focused on the SHG and SFG processes, the same device can be used to withstand downconversion processes such as difference frequency generation \cite{Sahin2021} and spontaneous parametric downconversion \cite{Dalidet2022}, or embedded in more complex photonic systems such as hybrid self-injection-locked sources \cite{Clementi2023achipscale}.
Finally, this approach is universal and could be extended to different material platforms, such as TFLN, in combination with an appropriate strategy to address the QPM condition \cite{Lu2020}.

In conclusion, we have demonstrated a frequency doubling device capable to arbitrarily address the doubly resonant condition, unlocking the potential for broadband and efficient upconversion.
Our approach can be extended to a variety of nonlinear integrated systems and processes, with applications ranging from microcombs to on-chip nonclassical light sources.

\vspace{0.5cm}
\noindent\textbf{Methods}
\medskip
\begin{footnotesize}

\vspace{0.1cm}

\noindent\textbf{Theoretical model}.
In order to estimate the  generated power, we can derive an equation for the SHG process. Considering the general expression for the second order nonlinear Hamiltonian, as given in \cite{Liscidini2012}, the fields of interest are those inside the upper arm of the Mach-Zehnder interferometer at the fundamental and second-harmonic frequencies and, if we explicit the terms in the equation, following the backward Heisenberg picture \cite{Yang2008} and consider the CW regime, we get for the power of the generated second harmonic:

\begin{equation}\label{eq:SHG_power}
    P_\text{SH} = \frac{ \left( \chi^{(2)}_\text{eff}P_\text{FH}\right)^2 \omega_\text{SH} \omega_\text{FH} \big|J(\omega_\text{SH},\omega_\text{FH},\omega_\text{FH}) \big|^2}{4 \varepsilon_0 n_\text{eff,SH}^2 n_\text{eff,FH}^4 v_\text{g,SH}v_\text{g,FH}^2 A_\text{eff}}  \ ,
\end{equation}
where $\chi^{(2)}_\text{eff}$ is the second order nonlinear susceptibility induced by the AOP, $P_\text{FH}$ is the injected pump power (in the bus waveguide), $n_{\text{eff}}$ is the effective index and $v_{\text{g}}$ is the group velocity (at the FH/SH frequency), $A_\text{eff}$ is the nonlinear effective area of the process, and $J(\omega_\text{SH},\omega_\text{FH},\omega_\text{FH})$ is the overlap integral, which takes into account the spatial overlap of the three fields involved in the process, that in our structure interact only inside the upper arm of the MZI of length $L_\text{MZI}$, and is given by:

\begin{align}\label{eq:overlap}
    &J(\omega_\text{SH},\omega_\text{FH},\omega_\text{FH}) = 
    f_\text{north}(\omega_\text{SH})f_\text{south}(\omega_\text{FH})^2 e^{i \frac{\Delta k L_\text{MZI}}{2}} \nonumber \\
    & \times   \text{sinc}\left(\frac{\Delta k L_\text{MZI}}{2}\right)\left[-\sigma_\text{SH} \kappa^2_\text{FH} + i \sigma_\text{FH}^2 \kappa_\text{SH}\right]L_\text{MZI} \ ,
\end{align}
where $f_\text{north}(\omega_\text{SH})$ is the field enhancement of the north resonator at the SH, while $f_\text{south}(\omega_\text{FH})$ is the field enhancement of the south resonator at the FH, $\sigma_\text{SH(FH)}$ and $\kappa_\text{SH(FH)}$ are the coupling coefficients of the directional couplers at the SH (FH), linked by the relation $\sigma_\text{SH(FH)}^2 + \kappa_\text{SH(FH)}^2 = 1$, and $\Delta k$ is the phase matching condition, which is:

\begin{equation}\label{eq:phase_matching}
    \Delta k = k_\text{SH}-2k_\text{FH} - \frac{2 \pi}{\Lambda} \ ,
\end{equation}
where $\Lambda$ is the poling period.
The resonant field enhancement inside the $i^{\text{th}}$ resonator can be expressed in term of quality factors as \cite{Menotti2019}:
\begin{equation}
    |f_i(\omega_0)|^2 = 
    \frac{4 v_{g,\omega_0}}{\omega_0L_i} \frac{Q_{\text{L},\omega_0}^2}{Q_{\text{C},\omega_0}} \ ,
\end{equation}
where $i=\text{north, south}$, $\omega_0=\omega_\text{SH},\omega_\text{FH}$ is the resonance frequency, $L_i$ is the length of the $i^{\text{th}}$ resonator, $Q_{\text{L},\omega_0}$ and $Q_{\text{C},\omega_0}$ are the loaded and coupling quality factors, respectively.
In our structure, considered linearly uncoupled by setting $\sigma_\text{SH}=\kappa_\text{FH}=1$, we have a nonlinear effective area $A_\text{eff}=\SI{1.34}{\micro m^2}$, loaded quality factors $Q_\text{L,SH}=\SI{1.78e6}{}$ and $Q_\text{L,FH}=\SI{0.69e6}{}$, coupling quality factors $Q_\text{C,SH}=\SI{9.69e6}{}$ and $Q_\text{C,FH}=\SI{3.05e6}{}$, total length of the resonators $L_\text{SH}=\SI{1129}{\micro m}$ and $L_\text{FH}=\SI{1124}{\micro m}$, and length of the Mach-Zehnder arm $L_\text{MZI}=\SI{361}{\micro m}$.
When the active AOP process sets the phase mismatch $\Delta k = 0$, inducing a $\chi^{(2)}_\text{eff} = \SI{0.1}{pm/V}$, we get for an injected power of $P_\text{FH}=\SI{300}{mW}$ a generated power of $P_\text{SH} \simeq \SI{20}{mW}$, with a conversion efficiency $\text{CE} \simeq 22\%/\si{\watt}$. 
 

\vspace{0.1cm}

\noindent\textbf{Experimental devices}.
The devices used in this work were fabricated through a commercially available foundry service (Ligentec SA).
The silicon nitride waveguides and resonators have a nominal cross-section of $1\times0.8\,\si{\micro\meter\squared}$ and they are embedded in a silica cladding.
Each fabricated photonic chip contains 69 devices (labeled “MZ1-MZ69”), all functional, designed to scan different values of north/south FSRs, to target different values of propagation loss in the two bands, and to compensate for fabrication imperfections.

After a first assessment of the linear properties of the devices fabricated, we focused our investigation on a single device (“MZ53”), which we used to obtain the results shown in this work.
The selected device displays a total south (north) loop length of \SI{1123.83}{\micro\meter} (\SI{1129.186}{\micro\meter}) for the FH (SH) optical path.
The south resonator is coupled to the bus waveguide through a point coupler with \SI{0.67}{\micro\meter} gap, while the north resonator relies on a \SI{62.1}{\micro\meter} long directional coupler with \SI{0.3}{\micro\meter} gap to efficiently in- and out-couple light to the bus waveguide, tailored for a target propagation loss of 0.2 dB/cm and 0.4 dB/cm for the FH/SH bands respectively.
Both resonators use Euler bends to mitigate mode mixing, with an effective bend radius around \SI{30}{\micro\meter}.

The MZI section consists of two \SI{32}{\micro\meter} long directional couplers with a gap of \SI{0.3}{\micro\meter}.
The upper arm of the MZI has a length of \SI{361.16}{\micro\meter} and the interferometer is unbalanced by \SI{2.197}{\micro\meter} to separately adjust the residual coupling of the FH and SH fields.

\vspace{0.1cm}

\noindent\textbf{Linear characterization}.
To characterize the linear properties of the device(s), we coupled light from a tunable laser (Toptica CTL at FH, New Focus Velocity TLB at SH) to the chip using a lensed fiber.
The sample contains inverse tapered waveguides, that provide an insertion loss between 2 and 3 dB per facet at the pump wavelength.
Light is collected in free space using an aspheric lens and routed either to InGaAs/Si detectors or coupled to fiber and routed to an optical spectrum analyzer.
Light at FH and SH is separated using a dichroic filter, while the input polarization is controlled to match the one of the fundamental TE mode.
To perform linear scans, we sweep the laser wavelength and record the transmitted power using a fast photodiode.
A fiber-based interferometer is used to calibrate the measurement to precisely assess the GVD.
The transmission spectra are recorded for varying power of the south, north, and MZI heaters (Supplementary Note 1).

\vspace{0.1cm}

\noindent\textbf{SHG investigation and optimization of the generated power}.
To investigate SHG, we proceed as follows.
On a selected device, we first tune the MZI heater to maximize visibility and Q factor in the pump band: this is done to prevent the leakage of pump power towards the north resonator resonances, that are  strongly overcoupled in the pump band and therefore act as a parasitic loss channel.
We then amplify the pump using an erbium-doped fiber amplifier (EDFA) to obtain a typical power around \SI{320}{\milli\watt} in the bus waveguide and tune it in resonance with the target south resonator mode.
We then scan the north ring resonances from lower to higher frequencies by slowly ramping down the current driving the north heater.
As soon as a nearly doubly resonant condition is matched, the onset for the AOP process occurs, highlighted by a sudden increase in the generated SH power, which settles to a stable value in a time ranging from few milliseconds to several seconds.
Finally, we slightly optimize the three heaters' currents to maximize the generated SH.

\vspace{0.1cm}

\noindent\textbf{TPM imaging}. 
For characterization of the inscribed $\chi^{(2)}$ gratings, a high power femtosecond Ti:Sapphire laser is focused at the grating plane of the microresonator in an upright configuration. 
The focal spot is then raster-scanned across the plane while, in the meantime, its generated SH signal is monitored so that the (squared) $\chi^{(2)}$ response is probed.
From the periodicity retrieved, the original phase mismatch between the modes involved is inferred as:
\begin{equation}
\frac{2\pi}{\Lambda}=
\frac{2\omega}{c}
\lvert n_{\rm eff}^{\rm SH} - n_{\rm eff}^{\rm FH}\rvert
\end{equation}
where $\Lambda$ is the poling period, $\omega$ is the pump angular frequency and $n_{\rm eff}^{\rm FH(SH)}$ is the effective index of the FH (SH) mode.
From simulations, we estimate $\Lambda=\SI{4.35}{\micro\meter}$, in good agreement with the result shown in Fig.~\ref{fig:fig2}d.


\end{footnotesize}

\vspace{0.5cm}
\noindent\textbf{Data availability}

\begin{footnotesize}
\noindent
The data and code that support the plots within this paper and other findings of this study are available from the corresponding authors upon request.

\end{footnotesize}

\bibliography{references}

\vspace{0.5cm}
\noindent\textbf{Acknowledgements}
\medskip
\begin{footnotesize}

\noindent
M.C., J.Z., and C.-S.B. acknowledge funding by the European Research Council grant PISSARRO (ERC-2017-CoG 771647) and by the Swiss National Science Foundation (grant 214889).
M.L. acknowledges PNRR MUR project “National Quantum Science and Technology Institute” – NQSTI (Grant No. PE0000023).

\end{footnotesize}

\vspace{0.5cm}
\noindent\textbf{Author contributions}

\begin{footnotesize}
\noindent
M.C. led the project, designed the devices, performed the measurements, and carried out the analysis and physical interpretation of the results.
L.Z. and M.L. contributed to the device design and theoretical modeling.
J.Z. performed preliminary characterization of the experimental devices.
M.C., M.L., and C.-S.B. conceived the original idea.
All authors participated in the analysis and discussion of the results.
M.C. wrote the manuscript with assistance from C.-S.B. and contributions from all authors.
C.-S.B. supervised the project.

\end{footnotesize}

\vspace{0.5cm}
\noindent\textbf{Competing interests}

\begin{footnotesize}
\noindent
The authors declare no competing interests.

\end{footnotesize}

\end{document}


\title{Supplementary Information for: \\Ultrabroadband Milliwatt-Level Resonant Frequency Doubling on a Photonic Chip}

\author{Marco Clementi\orcidlink{0000-0003-4034-4337}}
\thanks{Now at: Dipartimento di Fisica “A. Volta”, Università di Pavia, Via A. Bassi 6, 27100 Pavia, Italy}
\email{marco.clementi01@unipv.it}
\affiliation{Photonic Systems Laboratory, \'{E}cole Polytechnique F\'{e}d\'{e}rale de Lausanne,  1015 Lausanne, Switzerland}
\author{Luca Zatti\orcidlink{0000-0002-1280-9400}}
\affiliation{Dipartimento di Fisica “A. Volta”, Università di Pavia, Via A. Bassi 6, 27100 Pavia, Italy}
\author{Ji Zhou\orcidlink{0000-0001-8044-4426}}
\affiliation{Photonic Systems Laboratory, \'{E}cole Polytechnique F\'{e}d\'{e}rale de Lausanne,  1015 Lausanne, Switzerland}
\author{Marco Liscidini\orcidlink{0000-0003-4001-9569}}
\affiliation{Dipartimento di Fisica “A. Volta”, Università di Pavia, Via A. Bassi 6, 27100 Pavia, Italy}
\author{Camille-Sophie Brès\orcidlink{0000-0003-2804-1675}}
\affiliation{Photonic Systems Laboratory, \'{E}cole Polytechnique F\'{e}d\'{e}rale de Lausanne,  1015 Lausanne, Switzerland}

\date{\today}

\maketitle

\newpage

\section{Supplementary Note 1: Effect of the MZI tuning}
\begin{figure*}[ht!]
    \centering
    \includegraphics{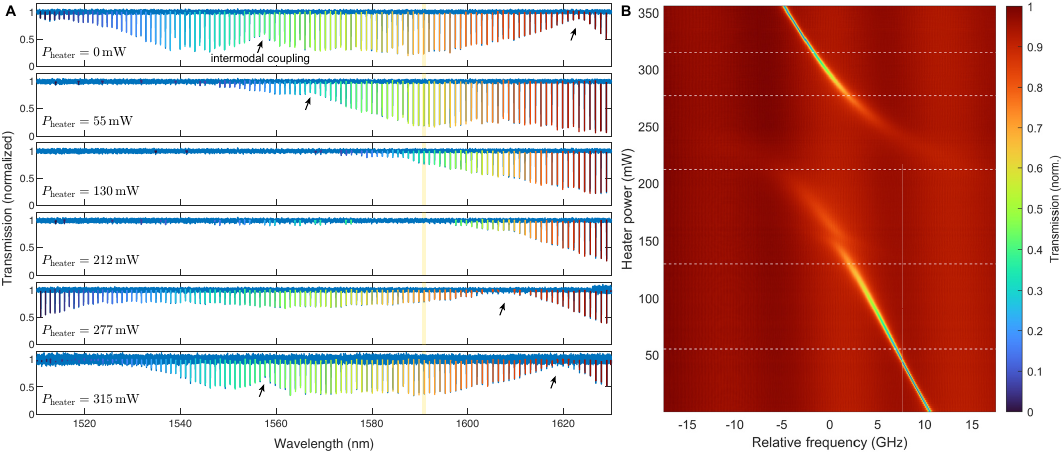}
    \caption{
    \textbf{Optimization of the linear uncoupling through MZI tuning.}
    \textbf{a.} Linear transmission spectra for increasing values of the MZI heater power.
    The maximum resonance visibility is obtained when the linear coupling is minimized.
    The features highlighted by black arrows are attributed to intermodal coupling.
    \textbf{b.} High resolution transmission spectra for a resonance around \SI{1590.9}{\nano\meter} (shaded area in panel \textbf{a}).
    The highest Q and visibility are obtained for a heater power of either \SI{35}{\milli\watt} or \SI{340}{\milli\watt}.
    Horizontal dashed lines mark the heater powers shown in panel \textbf{a}.
    }
    \label{fig:supp_mzi_resonances}
\end{figure*}

\noindent
Here we discuss the effect of residual linear coupling between the north and south resonators and the role of the MZI heater.
Supplementary Figure \ref{fig:supp_mzi_resonances}a shows the linear transmission spectra from the bus waveguide of the south resonator in the pump band for varying values of the MZI heater power.
Ideally, in the complete absence of linear coupling the spectrum should display a set of nearly critically coupled resonances.
Fabrication imperfections and the wavelength dependence of the directional couplers composing the MZI structure make it deviate from the ideal 100:0 splitting ratio, introducing a small residual linear coupling in the pump band.
This manifests as a lower resonance visibility and broader linewidth in some regions of the transmission spectrum: the north resonator is indeed strongly overcoupled in the pump band, and therefore acts as a loss channel whenever some residual linear coupling exists.
Such leak of power towards the north resonator can also be observed through collection from the corresponding (north) bus waveguide.
We mitigate this issue by acting on the MZI heater, which can be tuned to reach nearly ideal linear uncoupling in the desired spectral region, as shown in Supplementary Figure \ref{fig:supp_mzi_resonances}a.
In Supplementary Figure \ref{fig:supp_mzi_resonances}b, a high resolution scan focusing on a single resonance around \SI{1591}{\nano\meter} reveals, as an example, optimal visibility and narrowest linewidth for a heater power around 35 mW.
For higher powers, the resonance undergoes a linewidth broadening, while the narrow linewidth is recovered for a heater power around 340 mW.
The resonance also displays an anticrossing-type behavior, which we attribute to the hybridization with the north resonator low-Q TE$_{00}$ resonance.
Finally, we note that the transmission spectra also display local features associated with lower visibility and broader linewidth (black arrows).
We attribute these to weak intermodal coupling between the south resonator TE$_{00}$ mode and other mode families in the north resonator.

\section{Supplementary Note 2: Mapping of the AOP process}
\label{sec:supp_mapping}

\begin{figure*}[ht!]
    \centering
    \includegraphics{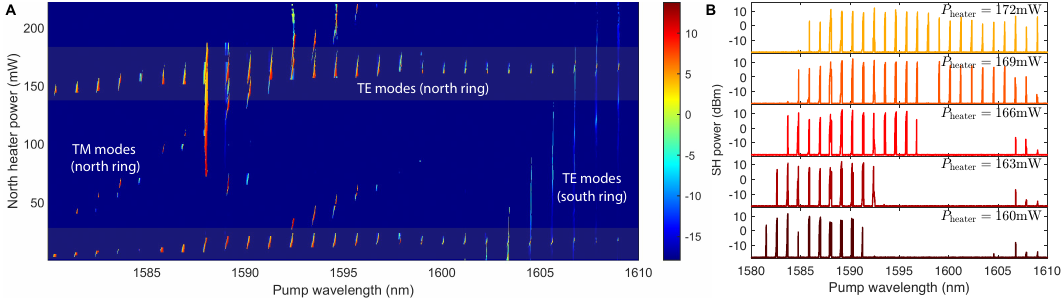}
    \caption{
    \textbf{Reconfiguration of the nonlinear interaction.}
    \textbf{a.} SHG map as a function of pump wavelength and north heater power.
    The trends highlighted correspond to doubly resonant SHG involving the TE$_{00}$-TE$_{00}$ mode pair.
    \textbf{b.} Detail from panel \textbf{a} showing typical SHG spectra for various values of the pump north heater power.
    }
    \label{fig:supp_mapping}
\end{figure*}

\noindent
To gather insight on the properties of the developed device, we carry out an extensive mapping of the AOP process across the L band \cite{Nitiss2023, Clementi2023achipscale}.
By slowly (\SI{100}{\pico\meter\per\second}) scanning the pump wavelength while keeping constant its power, we record the generated SH power for different driving conditions of the north heater.
The results, shown in Supplementary Figure~\ref{fig:supp_mapping}a, highlight the presence of several mode families at the SH, that can be visually identified as linear trends in the two-dimensional map.
In particular, the target mode family (i.e. the SH TE$_{00}$ modes of the north resonator) are represented by horizontal trends (shaded regions).
This is expected since the FSR of this mode family at the pump is matched with the one of the north ring at the SH: the result is that, for the correct value of north heater power, efficient SHG can occur across multiple resonances at the time.
Examples of SHG spectra for this mode family are shown in Supplementary Figure~\ref{fig:supp_mapping}b.
Other linear trends are visible on the AOP map: we identify a first one associated with the AOP of TM$_{00}$ modes in the north resonator, and a second one with AOP of the TE$_{00}$ modes of the south ring, as revealed by polarization analysis of the generated SH in combination with imaging of the device during the AOP process.
Occasionally, the SHG process involving other resonances than the target TE$_{00}$ modes of the north resonators (or a cascaded SHG-SFG process resulting in the generation of green light \cite{Hu2022}) can trigger a competition  in the AOP process.
However, this regime can generally be avoided by appropriate tuning of the heaters.

\section{Supplementary Note 3: Higher order stimulated FWM comb}
\begin{figure*}[ht!]
    \centering
    \includegraphics{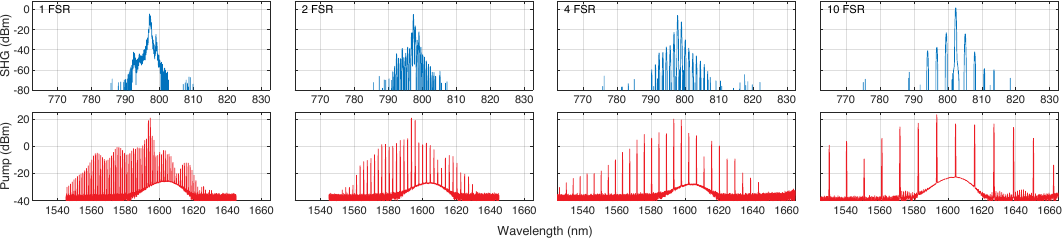}
    \caption{
    \textbf{Multi-FSR comb.}
    Dependence of the stimulated FWM comb on the pump lines spacing, as a multiple of the south resonator FSR.
    }
    \label{fig:supp_multi_fsr}
\end{figure*}

\noindent
Supplementary Figure~\ref{fig:supp_multi_fsr} shows the evolution of the stimulated FWM comb and its upconverted spectrum for varying values of the FSR.
Here, a dual-tone pump is used as shown in Fig. 4 of the main text, while the pump lines spacing is tuned to an integer multiple of the south resonator's FSR.
It can be noted that, while the pump comb extent and pump-to-line CE increases for increasing values of the spacing, the extend of the upconverted spectrum barely changes.
We attribute this phenomenological observation to the residual mismatch between the south and north resonators FSR in the respective bands.

\section{Supplementary Note 4: Resonator dispersion}
\begin{figure*}[ht!]
    \centering
    \includegraphics{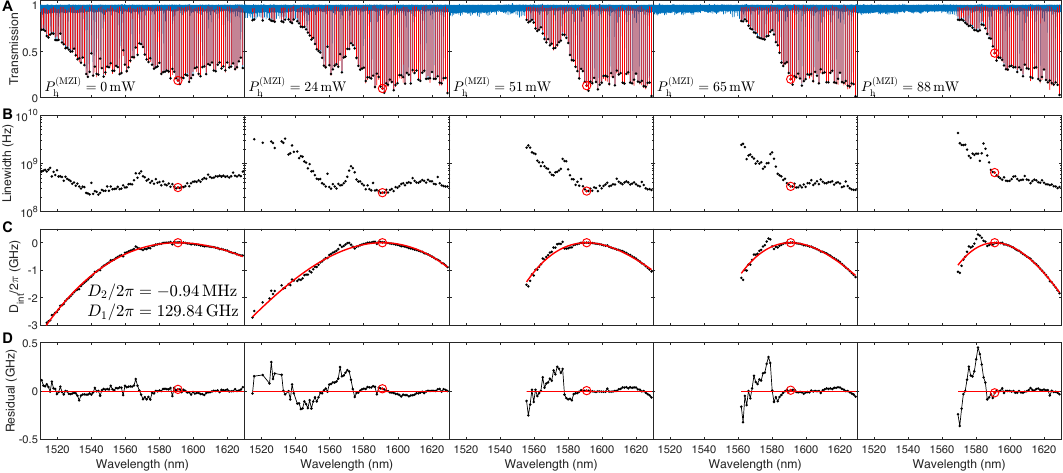}
    \caption{
    \textbf{Evolution of the resonator dispersion and linewidth for varying MZI heater power.}
    \textbf{a.} Low power, normalized transmission spectra.
    Red curves are Lorentzian fits of the recorded resonances and black dots indicate the minimum transmission.
    The red dot marks the pump resonance used for Kerr comb generation, around 1591~nm.
    \textbf{b.} Linewidth (full-width at half-maximum) values retrieved from best fit.
    \textbf{c.} Integrated dispersion profile.
    The red curve is a fit of the recorded data up to third order according with Eq. \eqref{eq:dint}.
    \textbf{d.} Residual from best fit of data in panel \textbf{c}.
    Note the intensification and contextual redshift of the perturbation (associated with broader linewidth and shifted GVD profile) for increasing heater power.
    }
    \label{fig:supp_dispersion}
\end{figure*}

\noindent
Here we investigate the formation mechanism for the comb states shown in Fig. 5 of the main text. 
Supplementary Figure~\ref{fig:supp_dispersion} shows transmission spectra for varying values of the MZI heater power. 
The laser wavelength was calibrated using a SMF28 fiber loop of known length and GVD.
First, we note that the resonances visibility and linewidth are strongly affected by the tuning of the MZI heater (Supplementary Figure~\ref{fig:supp_dispersion}a-b).
For a heater power comparable with that used to generate the MI comb (around 65 mW, fourth column), a high visibility is achieved for the pump resonance, while resonances at lower wavelength display broader linewidth.
This narrow wavelength dependence of the visibility relaxes the conditions required for the pump comb generation, while not significantly affecting the threshold.
A second factor affecting the comb formation can be deduced from the measurements of integrated dispersion relative to the pump resonance frequency (Supplementary Figure~\ref{fig:supp_dispersion}c), defined as:
\begin{equation}
    D_\mathrm{int}(\mu) =
    \omega_\mu - \omega_0 - D_1\mu = 
    \sum_{k\geq2} \frac{D_k}{k!} \mu^k
    \label{eq:dint}
\end{equation}
where $\omega_\mu/2\pi$ is the frequency of the $\mu$-th order resonance with respect to the pump, $D_1/2\pi = 129.84$ GHz is the microresonator’s FSR, and $D_2/2\pi = -0.94$ MHz is the group-velocity dispersion parameter, as determined from best fit of data at low heater power.
This indicates that the resonator is in normal GVD regime.
On the other hand, the presence of an anti-crossing type perturbation, also marked by a local increase of the resonances linewidth, alters the otherwise quadratic trend, facilitating the formation of MI combs \cite{Xue2016, Fujii2018}.
We note that the entity of this perturbation increases for increasing values of heater power, which we attribute to the increased linear coupling between the mode families of the resonators involved, while its spectral position gradually redshifts, increasing its effect on the pump resonance.
The optimal condition experimentally found for MI comb generation is a compromise between the resonance shift imparted by such resonance coupling and its detrimental effect on the pump threshold associated with linewidth broadening.

\section{Supplementary Note 5: Noise of the modulation instability comb}
\begin{figure*}[ht!]
    \centering
    \includegraphics[width=0.8\textwidth]{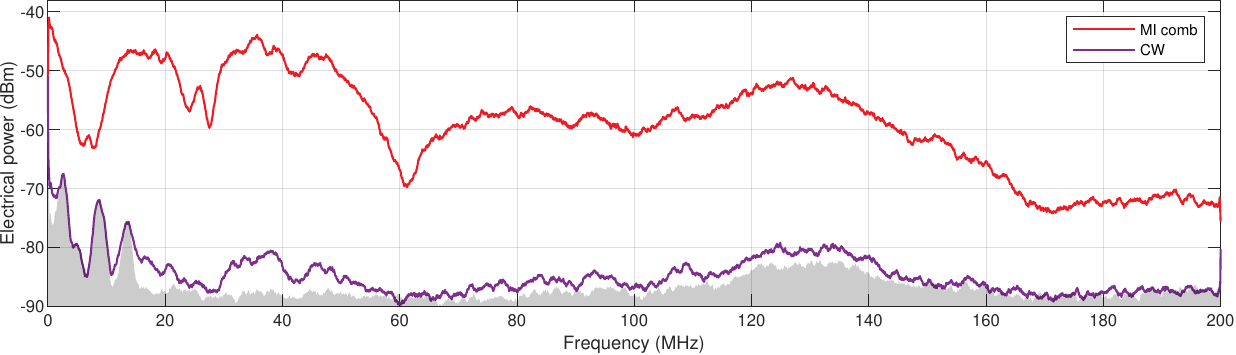}
    \caption{
    \textbf{Modulation instability comb noise spectrum.}
    Electrical spectrum analyzer spectra of a generated comb spectrum at FH (red) and of the pump out of resonance (purple) for the generated MI comb state shown in Fig.~5a of the main text.
    The gray shaded area represents the detector noise floor.
    }
    \label{fig:supp_esa}
\end{figure*}

\noindent
Supplementary Figure~\ref{fig:supp_esa} shows an electrical noise spectrum associated with the generated modulation instability comb shown in Fig.~5a of the main text.
Here, light in the FH band at the output of the south resonator is sent to a fast photodiode and the photodetected current is  routed to an electrical spectrum analyser (ESA), where the electrical noise spectrum is recorded.
Compared with the noise spectrum of the continuous wave (CW) comb pump alone, which is close to the detector's noise floor, the generated comb yields highly noisy spectrum, which is typical of MI states.
This observation confirms the incoherent nature of the generated comb.

\section{Supplementary Note 6: Extended data for Figure 2c}
\begin{figure*}[ht!]
    \centering
    \includegraphics[width=1\textwidth]{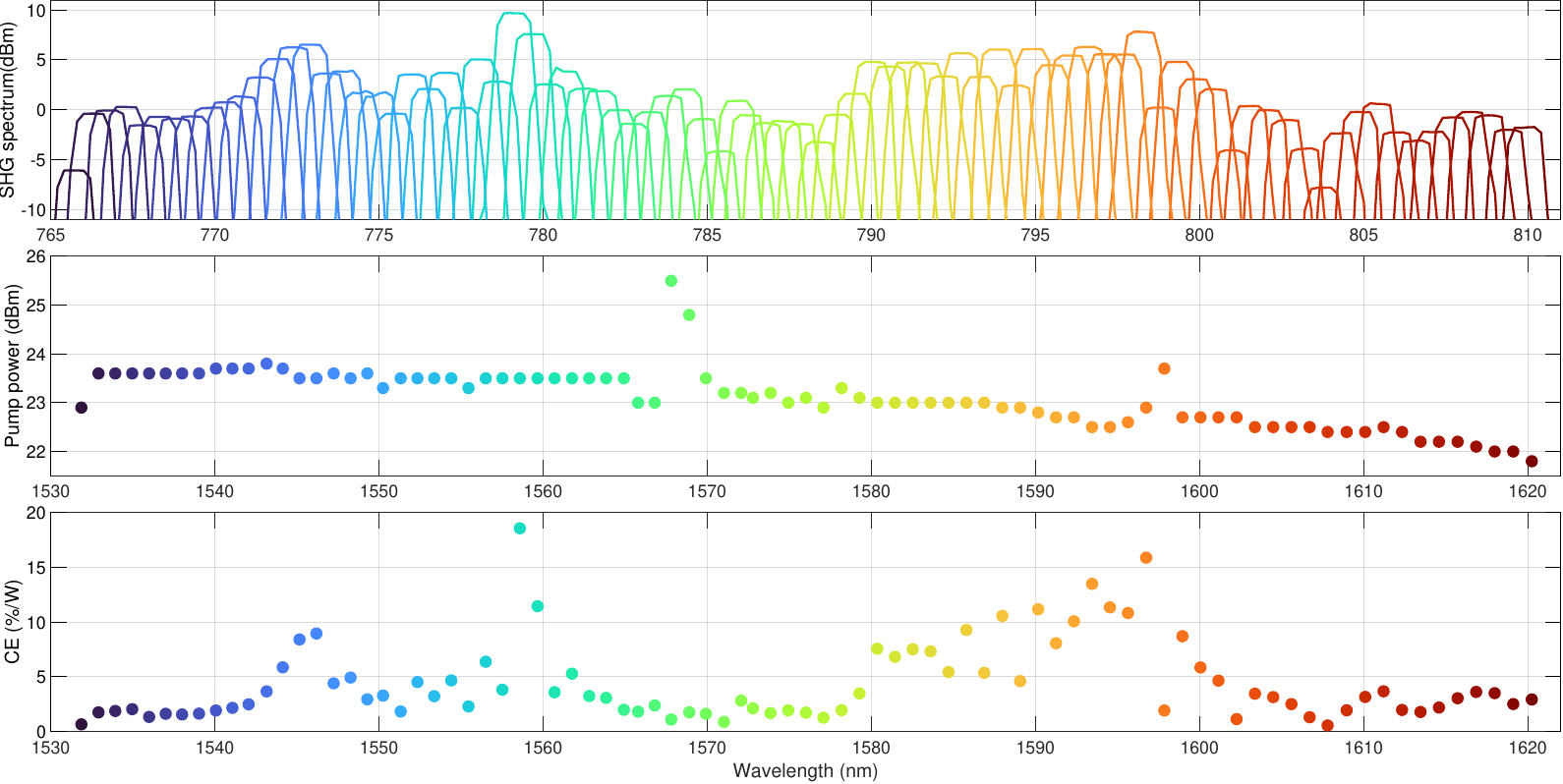}
    
    \caption{\textbf{Reconfigurable SHG spectra.}
    (\textit{upper panel}) Collection of SHG spectra associated with pumping in the C and L bands (resolution: \SI{1}{\nano\meter}).
    (\textit{mid and lower panels}) Adjusted pump power in the waveguide and the calculated conversion efficiency for each data point.
    The heater currents are slightly adjusted for each data point to optimize the generated power.
    Each trace and experimental point is a detailed version of those shown in Fig. 2c of the main text.
    }
    \label{fig:supp_extended}
\end{figure*}

\newpage

\bibliography{references}